\documentclass[useAMS,usenatbib]{mn2e}
\usepackage{graphicx}

\title[Discovery of a widely separated UCD-WD binary]
{Discovery of a widely separated ultracool dwarf -- white dwarf binary}
\author[A. C. Day-Jones et al.]{
A. C. Day-Jones$^1$\thanks{E-mail: a.c.day-jones@herts.ac.uk. Based on
observations made with ESO telescopes at the La Silla Paranal
Observatory under programme 278.C-5024(A)}, D. J. Pinfield$^1$,
R. Napiwotzki$^1$, B. Burningham$^1$, J. S. Jenkins$^{1,2}$,\\ \newauthor
H. R. A. Jones$^1$, S. L. Folkes $^1$, D. J. Weights$^1$ and J. R. A. Clarke$^1$\\ $^{1}$
Centre for Astrophysics Research, University of Hertfordshire,
College Lane, Hatfield, Hertfordshire, AL10 9AB, UK\\
$^{2}$
Penn State University, State College, PA 16802, U.S.}

\begin{document}

\date{}

\pagerange{\pageref{firstpage}--\pageref{lastpage}} \pubyear{2007}

\maketitle

\label{firstpage}

\begin{abstract}
We present the discovery of the widest known ultracool dwarf -- white
dwarf binary. This binary is the first spectroscopically confirmed
widely separated system from our target sample. We have used the 2MASS
and SuperCOSMOS archives in the southern hemisphere, searching for
very widely separated ultracool dwarf -- white dwarf dwarf binaries,
and find one common proper motion system, with a separation of
$3650-5250$\,AU at an estimated distance of $41-59$\,pc, making it the
widest known system of this type.  Spectroscopy reveals 2MASS
J$0030-3740$ is a DA white dwarf with $T_{\rm eff}$=$7600\pm100$K,
log($g$)$=7.79-8.09$ and M$_{WD}=0.48-0.65$\,M$_{\odot}$. We
spectroscopically type the ultracool dwarf companion (2MASS
J$0030-3739$) as M9$\pm$1 and estimate a mass of
$0.07-0.08$\,M$_{\odot}$, $T_{\rm eff}=2000-2400$\,K and
log($g$)=$5.30-5.35$, placing it near the mass limit for brown
dwarfs. We estimate the age of the system to be $>$1.94\,Gyrs (from
the white dwarf cooling age and the likely length of the main sequence
lifetime of the progenitor) and suggest that this system and other
such wide binaries can be used as benchmark ultracool dwarfs.

\end{abstract}


\section{Introduction}
The rise of deep large area surveys such as 2MASS (the Two Micron All
Sky Survey), SDSS (the Sloan Digital Sky Survey) and UKIDSS (the UK
Infrared Deep Sky Survey) has increased the number of known brown
dwarfs to several hundred, since the discovery of Gliese 229B
\citep{nakajima95} and Tiede 1 \citep{rebolo95} just over a decade
ago. These populations have helped shape our understanding of
ultracool dwarfs (UCDs; with spectral type $\ge$M7 e.g.
\citealt{jones01}) and revolutionized the classification system for
sub-stellar objects including the creation of two new spectral types L
and T. The latest M dwarfs ($\sim$M7-9) have effective temperature
($T_{\rm eff}$) reaching down to $\sim$2300K. At lower $T_{\rm eff}$
($\sim$2300-1400\,K) are the L dwarfs, which have very dusty upper
atmospheres and generally very red colours. T dwarfs are even cooler
having $T_{\rm eff}$ in the range $\sim$1400-650\,K, where the low
$T_{\rm eff}$ limit is currently defined by the recently discovered
T8$+$ dwarfs, ULAS J0034-0052 \citep{warren07} and CFBDS
J005910.90-011401.3 \citep{delorme08}. T dwarf spectra are dominated
by strong water vapour and methane bands, and generally appear very
blue in the near infrared (\citealt{geballe03};
\citealt{burgasser06}).

The physics of ultracool atmospheres is complex, and very difficult to
accurately model. Atmospheric dust formation is particularly
challenging for theory (\citealt{allard01}; \citealt{burrows06}), and
there are a variety of other important issues that are not well
understood, including the completeness of CH${_4}$/H${_2}$O molecular
opacities, their dependence on $T_{\rm eff}$, gravity and metallicity
(e.g. \citealt{jones05}; \citealt{burgasser06a}; \citealt{liu07}), as
well as the possible presence of vertical mixing in such atmospheres
\citep{saumon07}. The emergent spectra from ultracool atmospheres are
strongly affected by factors such as gravity and metallicity (e.g.
\citealt{knapp04}; \citealt{burgasser06a}; \citealt{metchev06}), and
we must improve our understanding of such effects if we are to be able
to spectroscopically constrain atmospheric and physical properties
(such as mass, age and composition) of low-mass/sub-stellar field
populations.

Discovering UCDs whose properties can be inferred indirectly (without
the need for atmospheric models) is an excellent way to provide a
test-bed for theory, and observationally pin down how physical
properties affect spectra. We refer to such UCDs as {\it benchmark}
objects (e.g. \citealt{pinfield06}). A population of benchmark UCDs
with a broad range of atmospheric properties will be invaluable in the
task of determining the full extent of spectral sensitivity to
variations in UCD physical properties. However, such benchmarks are
not common, and the constraints on their properties are not always
particularly strong.

One variety of benchmark UCD that could yield accurate ages and
surface gravities over a broad age range are in binary systems with a
white dwarf (WD) companion. In particular, if the WD is relatively
high mass, then the main sequence progenitor star will have a short
lifetime, and the age of the binary system (and the UCD) will
essentially be the same as the cooling age of the WD, which can be
well determined from theory.

There have been several searches to find UCD companions to
WDs. Despite this, only a small number of detached UCD--WD binaries
have been identified; GD 165B(L4) \citep{zuckerman92}, GD 1400(L6/7)
(\citealt{farihi04}; \citealt{dobbie05}), WD\,$0137-349$(L8)
(\citealt{maxted06}; \citealt{burleigh06}) and PG1234+482 (L0)
(\citealt{steele07}; \citealt{mullally07}).  The two components in
GD\,165 are separated by 120\,AU, the separation of the components in
GD\,1400 and PG\,$1234+482$ are currently unknown and WD\,$0137-349$
is a close binary (semi-major axis
$a=0.65$\,R$_\odot$). \citet{farihi05} and \citet{farihi06} also
identified three late M companions to white dwarfs; WD$2151-015$ (M8
at 23\,AU), WD$2351-335$ (M8 at 2054\,AU) and WD$1241-010$ (M9 at
284\,AU). The widest system previously known was an M8.5 dwarf in a
triple system -- a wide companion to the M4/WD binary LHS 4039 and LHS
4040 \citep{scholz04}, with a separation of 2200AU. There are several
other known UCD--WD binaries, however these are cataclysmic variables
(e.g. SDSS 1035; \citealt{littlefair06}, SDSS1212;
\citealt{burleigh06b}, \citealt{farihi08}, EF Eri; \citealt{howell01})
but these are unlikely to provide the type of information that will be
useful as benchmarks, as they have either evolved to low masses via
mass transfer or their ages cannot be determined because of previous
evolutionary phases (e.g. common envelope). WD0137-349 is also not
likely to be useful as a benchmark due to the large uncertainty in age
caused by the common envelope stage and the heating from the white
dwarf.

Recent analysis from \citet{farihi08b} shows that the fraction of
brown dwarf companions (L and T) at separations within a few hundred
AU of white dwarfs is $<$0.6 per cent.  However, despite this UCDs in
wide binary systems are not uncommon (revealed through common proper
motion) around main sequence stars at separations of $1000-5000$\,AU
(\citealt{gizis01}; \citealt{pinfield06}). However, when a star sheds
its envelope post-main sequence, we may expect a UCD companion to
migrate outwards to even wider separation (\citealt{jeans1924};
\citealt{burleigh02}), and UCD--WD binaries could thus have
separations up to a few tens of thousands of AU. Although some of the
widest binaries may be dynamically broken apart quite rapidly by
gravitational interactions with neighbouring stars, some systems may
survive, offering a significant repository of benchmark UCDs.

We present here the first results from our search of 2MASS and the
digitized Schmidt plate archive SuperCOSMOS, for wide UCD--WD binaries
in the southern sky, and present our first discovery of a wide UCD--WD
binary confirmed through common proper motion and spectroscopy. In
Section 2 we describe the techniques we have developed to select
candidate UCD--WD binary systems using database photometry and proper
motion. Section 3 describes our second epoch imaging of candidates,
and followup infrared and optical spectroscopy along with proper
motion analysis. We present spectroscopic analysis in Section
4. Section 5 discusses the newly discovered system, and Section 6
presents a discussion of future work.

\section{Candidate Binary Selection}
\subsection{Selecting white dwarf candidates}
\begin{figure*}
\includegraphics[width=135mm, angle=90]{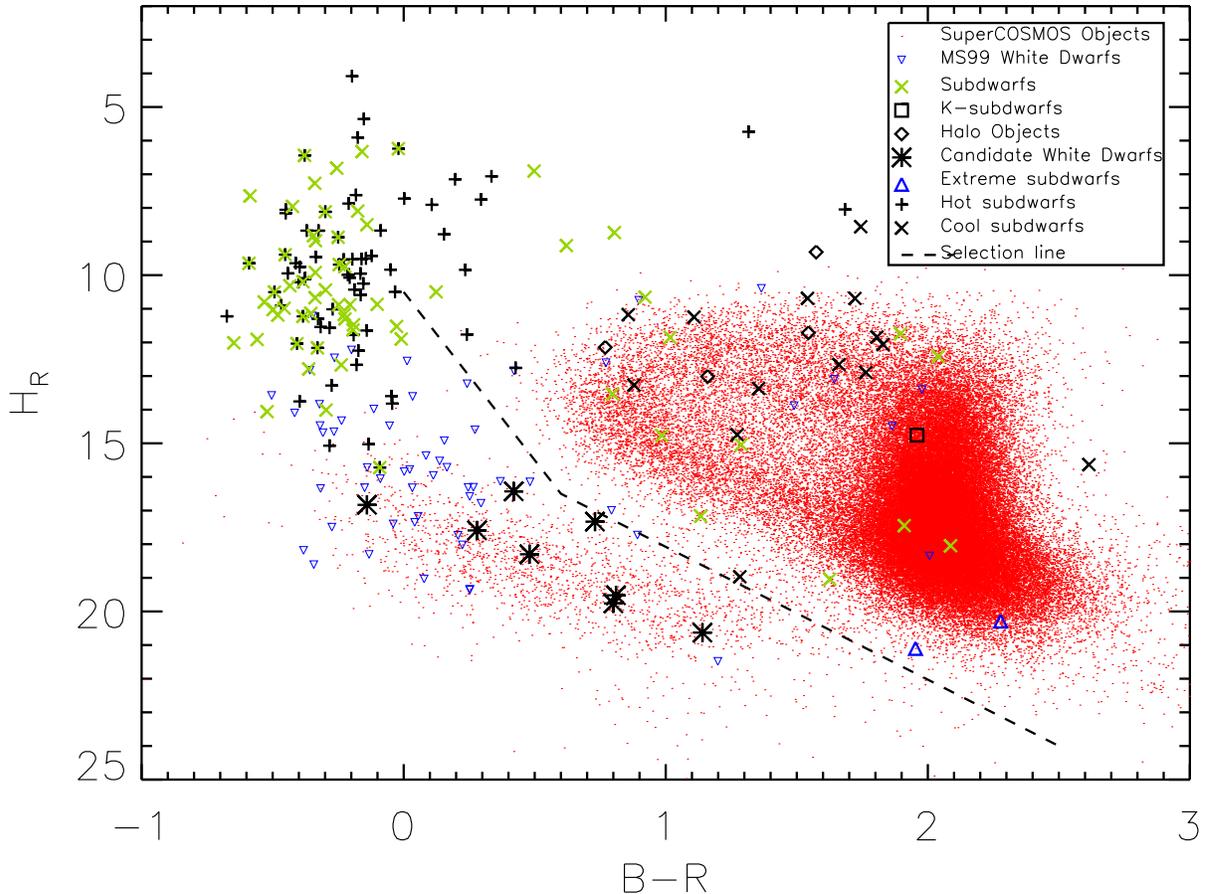}
\caption{Reduced Proper Motion (RPM) diagram showing how WD candidates
were selected from SuperCOSMOS. We chose candidate WDs from amongst
our initial (see text) SuperCOSMOS sample (points) using a cut in
colour-RPM space (RPM=H${_R}=R+5log\mu +5$), which is overplotted in
the figure as a dashed line. Highlighted are spectroscopically
confirmed WDs from MS99 (upside down triangles), hot subdwarfs (plus
signs), cool subdwarfs (crosses) and extreme subdwarfs (triangles) to
help delineate the WD sequence. The location of some halo objects
(diamonds and squares) are also indicated. Also shown are the WD
components of our eight candidate UCD -- WD binaries (stars).}
\label{rpmd}
\end{figure*}

Candidate WDs were selected from the SuperCOSMOS Science Archive
(SSA), following a similar technique to that of \citet{knox99}. We
selected candidates based on their position on a reduced proper motion
(RPM) diagram, a technique that uses proper motion as a distance
proxy, taking advantage of the fact that nearby objects in general
have higher proper motion. Fig.~\ref{rpmd} shows our SuperCOSMOS RPM
diagram. Our initial SSA sample (shown as dots) was a magnitude
selected sample ($R\le$20) to avoid sources near the plate limit. We
also selected against non-moving sources (by requiring PM$\ge$10\,mas
yr$^{-1}$, PM/$\sigma_{PM}\ge$5), and against non stellar sources
(requiring a database star-like class=2). In addition we avoid the
galactic plane (requiring $|b|\ge$25$^{\circ}$), require SSA $I$- band
coverage ($\delta\le$+3$^{\circ}$; which offers possible additional
epochs for nearby UCD candidates; see Section 3.3), and impose a
quality constraint on the database photometry (qual$\le$1040 in each
of the $B$-, $R$- and $I$- bands, indicating the the object is
unlikely to be a bright star artifact.).

The main sequence and WD sequence can be clearly seen in
Fig.~\ref{rpmd}, and display good separation for $B-R\le$1.3. To
further illustrate the location of the WD sequence, we over plot the
positions of WDs from the spectroscopically confirmed McCook \& Sion
WD catalogue (\citealt{mcs99}; here after MS99), along with pre-WDs
(i.e. sdO stars) and Halo objects from \citet{leggett92}. We also
overplot subdwarf B stars, hot, cool and extreme subdwarfs from
\citet{kilkenny88}, \citet{stark03}, \citet{yong03} and \citet{monet92}
respectively. These, like halo objects can have high velocity
dispersions and can masquerade as white dwarfs in RPM diagrams. The
overplotted populations help confirm the location of the WD sequence
for $B-R\le$1.3, and allow us to assess the location of WDs out to
redder colour. Our final RPM selection criteria were chosen to strike
the best balance between WD selection and contamination minimisation,
and are shown in Fig.~\ref{rpmd} as dashed lines. At the red end in
particular, we have attempted to include as many WD candidates as
possible, while avoiding contamination from cool subdwarfs.

The RPM selection criteria are
\begin{center}
$H_{R}\ge{8.9(B-R)+10.5}$ for ${B-R}\le{0.65}$  and \\
$H_{R}\ge{4.1(B-R)+12.5}$ for ${B-R}>{0.65}$;
\end{center}

Our selection criteria resulted in a sample of 1532 WD candidates.

\subsection{Selecting ultracool dwarf candidates}
UCD candidates were photometrically selected by their near infrared
(NIR) colours from the 2MASS all sky point source catalogue, using the
NASA/IPAC Gator search engine. We used colour cut criteria devised by
\citet{folkes07}, which are based on the colours of known L-dwarfs
with reliable $J$-\,, $H$- and $K$- band 2MASS photometry (SNR$\ge$20) from the
Caltech cool dwarf archive ({\sc dwarfarchives.org}), requiring the
following photometric criteria to be met:
\begin{center}
$0.5\le{J-H}\le{1.6}$\\ $1.1\le{J-K}\le{2.8}$\\
$0.4\le{H-K}\le{1.1}$ \\ ${J-H}\le{1.75(H-K)+0.37}$\\
${J-H}\ge{1.65(H-K)-0.35}$\\ and $J\le{16.0}$. \\
\end{center}

We also imposed an optical-NIR colour restriction using the USNO-A2.0
cross-match facility within the Gator, ruling out objects with
$R-K<$5.5 ($a=U$ combined with vr\_m\_opt-$K>$5.5, or
nopt\_mchs=0). In addition certain sky coverage constraints were
imposed, and we made use of the 2MASS data quality criteria and some
additional database flags, to remove contamination from artefacts and
low quality photometric data (requiring cc\_flag=000,
ph\_qual$\le$CCC, jhk\_snr$>$5, gal\_contam=0, prox$>$6) and known
contaminants (mp\_flg=0). As with the WD candidate selection, we
searched areas away from the galactic plane ($|b|>$25$^{\circ}$) to
avoid confusion from over crowding and contamination from reddened
stars and giants. We also avoided $|\delta|>$86$^{\circ}$, for which
2MASS suffers from incompleteness issues with its optical
cross-matching. Additional areas surrounding the Small and Large
Magellanic clouds were also ignored. Finally two additional
uncatalogued, reddened regions were avoided following the approach of
\citet{cruz03}. These excluded regions are listed in
Table~\ref{cloud_table}, and the resultant sky coverage of our survey
is 13,216 sq. degs, or 32 per cent of the sky.  Visual inspection of
all these candidates (13,338 in number) was neither practical or
necessary at this stage, and we thus postponed our visual inspection
until after our binary pair selection procedure (see next Section).

\begin{table}
\caption{Galactic co-ordinates of contaminated and overcrowded regions
removed from our selection.}  \centering
\begin{tabular}{|l|c|c|c|c|}
\hline
Region & $l_{\rm min}$ & $l_{\rm max}$ & $b_{\rm min}$ & $b_{\rm max}$ \\
\hline
SMC.........& 300$^\circ$ & 310$^\circ$ & -42$^\circ$ & -23$^\circ$ \\
LMC.........& 270$^\circ$ & 290$^\circ$ & -40$^\circ$ & -25$^\circ$ \\
.................& 145$^\circ$ & 220$^\circ$ & -42$^\circ$ & -20$^\circ$ \\
.................&  20$^\circ$ &  45$^\circ$ & -25$^\circ$ & -20$^\circ$ \\
\hline
\end{tabular}
\label{cloud_table}
\end{table}

\subsection{Selection of candidate binary pairs}
To identify very widely separated UCD--WD binary candidates, we
searched for pairs with separations out to 20,000AU, allowing for an
outward migration factor of $\sim$4 (during the post-main-sequence
mass-loss phase) from the known separations of wide UCD --
main-sequence star binaries (5,000AU; see Section 1). To illustrate
this choice, consider a WD of mass $\sim$0.65\,M$_{\odot}$ (the mean of the
WD mass distribution). The progenitor mass would be $\sim$2.7\,M$_{\odot}$
(from the initial-final mass relation from \citealt{dobbie06}) so
M$_{initial}$/M$_{final}$ $\sim$ 4 \citep{jeans1924}. We thus suggest
that the projected maximum orbital separation could be up to $\sim$
20,000AU. In addition, we imposed the constraint that the photometry
of any UCD and WD pairs associated as candidate binaries, be
consistent with these objects being at the same distance. In order to
do this, reasonable distance constraints were required for both types
of object.

\begin{figure*}
\includegraphics[width=100mm, angle=90]{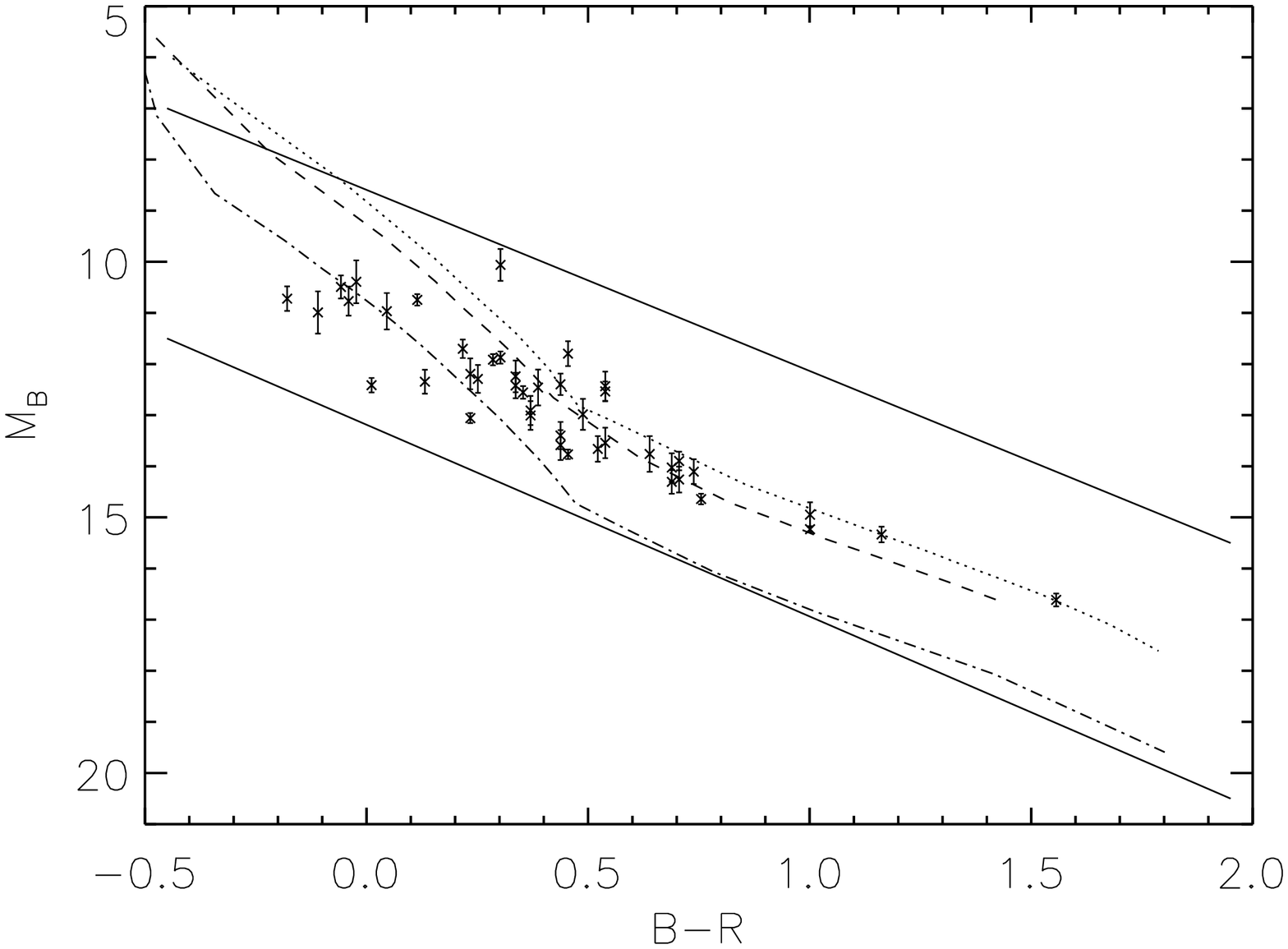}
\caption{WD colour-magnitude diagram for MS99 WDs with known parallax
(crosses).  Photometry is on the SuperCOSMOS system. Overplotted are
model cooling tracks (see main text) for WD masses of 0.5, 0.7 and
1.2\,M$_{\odot}$ (dotted, dashed and dot-dashed lines
respectively). Our WD region in the CMD lies between the two solid
lines.}
\label{brcmd}
\end{figure*}

We started by placing a distance constraint on the WDs, by
constructing an $M_B$ against $B-R$ colour magnitude diagram (CMD),
using a combination of observed and theoretical WD photometry; this
CMD is shown in Fig.~\ref{brcmd}. Known WDs were taken from MS99,
including all WDs with known parallax and distance uncertainties
better than 20 per cent. These objects are plotted as crosses with
associated error bars. In addition synthetic WD properties
(luminosity, $T_{\rm eff}$ and log($g$)) were obtained for three
different masses (0.5, 0.7 and 1.2\,M$_{\odot}$) over a range of disk
ages, using equations from \citet{schroeder04}, and the mass--radius
relation of \citet{panei00}. Synthetic WD photometric properties were
then determined using a combination of colour--$T_{\rm eff}$ and
BC--$T_{\rm eff}$ relations from models \citep{chabrier00} and
observation \citep{kleinmann04}. The photometry was transformed into
the SuperCOSMOS system using relations given in \citet{bessell86}.
These theoretical tracks are shown in Fig.~\ref{brcmd} as dotted,
dashed, and dot-dash lines respectively.

We defined a WD region in the CMD (shown by the solid lines) to take
into account the spread seen in both observation and the models, while
also offering a reasonably constrained WD sequence (that will yield
useful distance constraints). Although some of the hottest WD model
tracks lie slightly above our region, we note that no such trend is
seen for the observations, and that the highest mass hot model points
are completely contained within it. This is desirable since high mass
WDs are more interesting in the context of benchmark UCDs. Using this
CMD as an aid for characterising WDs, we used the $B-R$ colour for
each of our WD candidates to estimate a possible range of $M_B$, thus
derived a corresponding distance range appropriate to the measured
$B$- magnitude of each candidate.

\begin{figure}
\vspace{4.5cm}
\hspace{2.0cm}
\includegraphics{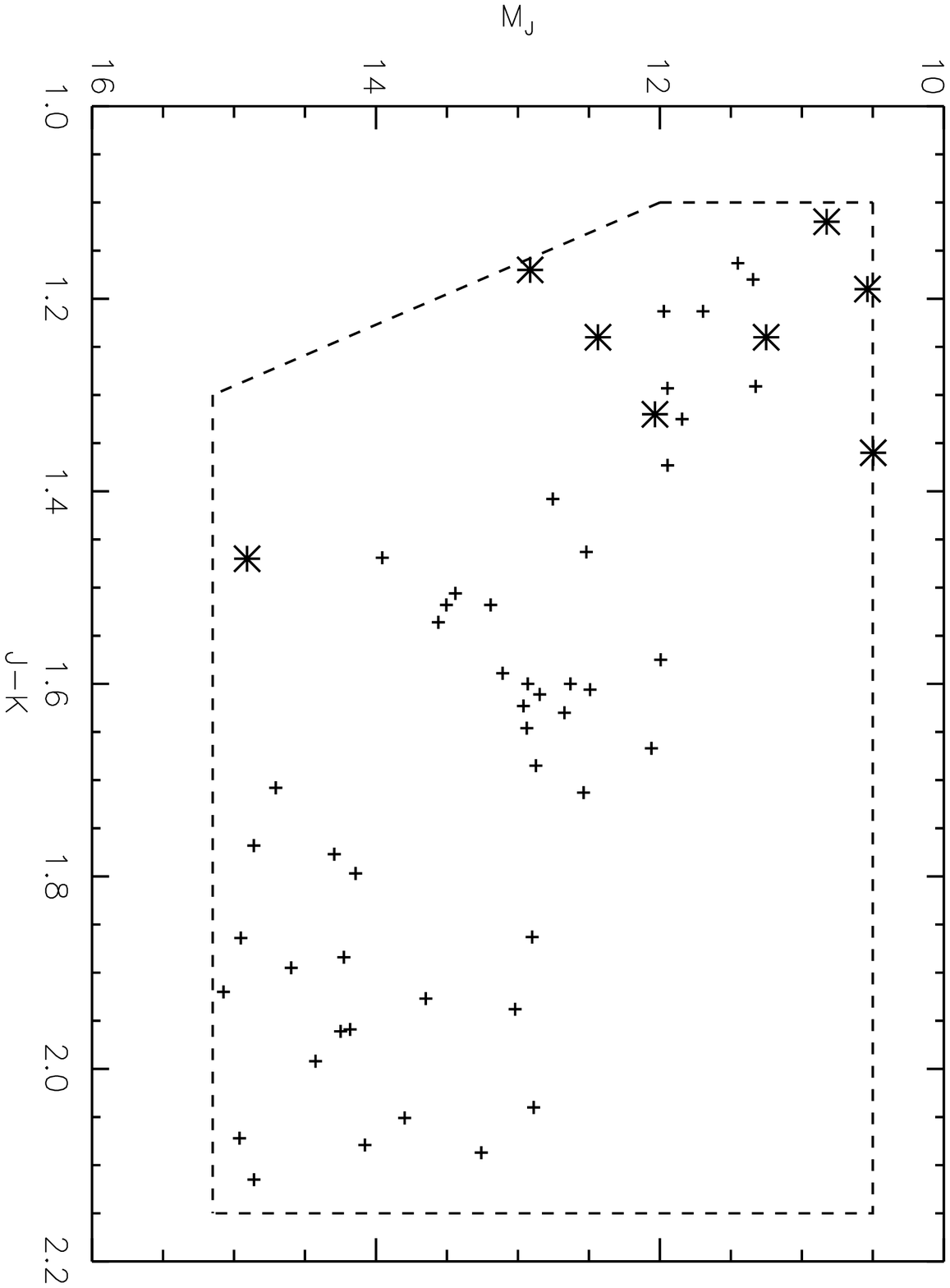}
\vspace{2.0cm}
\caption{$M_J$ against $J-K$ colour-magnitude diagram showing the location
of the companion UCD candidates (stars) when they are assumed to be at the
same distance as their associated WD. Our UCD selection criteria is indicated
with a dashed line, and UCDs with known parallax from {\sc dwarfarchives.org}
are plotted as plus signs.}
\label{ld_cmd}
\end{figure}

We then used the lower distance estimates to provide upper limits to
the angular separation corresponding to 20,000AU at the distance of
each of the WD candidates, and searched for UCD candidates whose
angular separation from the WDs was within the appropriate limit. We
also used the UCD colour-magnitude information to check for
consistency between candidate binary pairs, by using the WD distance
estimates to convert UCD candidate $J$- band magnitudes into
$M_J$. This assumes that the two objects are at the same distance. We
then plotted the UCD candidate on an $M_J$ against $J-K$ CMD, to see
if it was located in the expected part of the
diagram. Fig.~\ref{ld_cmd} shows this NIR CMD, with the known location
of previously confirmed UCDs (with parallax distances from Knapp et
al. 2004). We defined a similar UCD CMD selection region as
\citet{pinfield06}, which is shown as a dashed line box. If UCD
candidates lie outside this box, then their photometry is deemed
inconsistent with a UCD at the same distance as the neighbouring
candidate WD. All possible pairs were considered amongst our 1532 WD
and 13,338 UCD candidates, and assessed with our separation and
photometric consistency tests. In this way we identified 18 candidate
UCD--WD binary systems. These candidates were visually inspected using
images from 2MASS, SuperCOSMOS and DENIS (where available). We find
that seven of the UCD candidates have bright $R$- band counterparts
and were therefore rejected as their $R-J$ colour is too blue. Proper
motion analysis using $I$- band schmidt plates from SuperCOSMOS and
2MASS $J$- band images revealed that three are non common proper
motion. The UCD and WD components from the eight remaining candidate
pairs are presented in Table~\ref{photom_table} and over-plotted as
stars in Fig.~\ref{rpmd} and~\ref{ld_cmd}. The clumping in to the top
left of the selection space in Fig.~\ref{ld_cmd} reflects our greater
sensitivity to closer, brighter late M and L dwarfs in 2MASS.

\begin{table*}
\caption{Candidate UCD-WD binary systems. Coordinates are J2000. Photometry is
from 2MASS, SuperCOSMOS and DENIS (where available). The last three columns show which UCD candidates have
second epoch imaging (see Section 3.3), which pairs have been confirmed (or not) through common
proper motion and which candidates have been confirmed as a UCD or WD
with spectroscopy (see Section 4).}
\centering
\begin{tabular}{|l|l|l|c|c|c|c|c|c|l|c|c|c|}
\hline
Name & RA & Dec & $B$- & $R$- & $I$- & $J$- & $H$- & $K$- & Sep (arcsec) & 2nd epoch & CPM & Spec \\
\hline
UCDc-1 & 00 30 06.26 & -37 39 48.2 &     - &     - & 18.30 & 15.2 & 14.4 & 13.8 & 89  & IRIS2 & Y & Y \\
 WDc-1 & 00 30 11.90 & -37 40 47.2 & 16.77 & 16.35 & 15.97 & 16.1 & 15.8 &    - &     &       &   & Y \\
UCDc-2 & 00 56 14.71 & -40 36 03.6 &     - &     - &     - & 15.8 & 15.0 & 14.3 & 922 & IRIS2 & N & - \\
 WDc-2 & 00 57 18.72 & -40 45 29.6 & 15.80 & 15.07 & 13.67 & 12.4 &    - &    - &     &       &   & - \\
UCDc-3 & 03 02 07.70 & -09 41 57.1 &     - &     - & 17.66 & 15.8 & 15.0 & 14.5 & 136 & IRIS2 & N & - \\
 WDc-3 & 03 02 03.00 & -09 43 54.9 & 17.63 & 17.35 & 17.34 &    - &    - &    - &     &       &   & - \\
UCDc-4 & 05 20 35.40 & -18 54 27.7 &     - &     - & 17.64 & 15.9 & 15.1 & 14.7 & 145 & IRIS2 & N & - \\
 WDc-4 & 05 20 40.03 & -18 56 37.9 & 18.34 & 16.40 & 17.92 &    - &    - &    - &     &       &   & - \\
UCDc-5 & 10 12 35.59 & -10 51 02.2 &     - &     - & 18.13 & 15.3 & 14.6 & 14.1 & 465 & IRIS2 & ? & - \\
 WDc-5 & 10 12 43.89 & -10 43 33.8 & 19.48 & 18.34 & 17.87 &    - &    - &    - &     &       &   & - \\
UCDc-6 & 10 40 43.41 & -16 48 20.5 &     - &     - &     - & 15.9 & 15.2 & 14.8 & 124 & IRIS2 & ? & - \\
 WDc-6 & 10 40 39.17 & -16 50 08.3 & 20.34 & 19.54 & 19.12 &    - &    - &    - &     &       &   & - \\
UCDc-7 & 14 05 37.54 & -05 51 53.6 &     - &     - & 17.80 & 15.8 & 15.2 & 14.6 & 164 & - & - & - \\
 WDc-7 & 14 05 44.98 & -05 49 51.9 & 16.66 & 16.80 & 16.87 &    - &    - &    - &     &       &   & - \\
UCDc-8 & 23 21 21.55 & -13 26 28.3 &     - &     - &     - & 14.5 & 13.5 & 13.1 & 73 & IRIS2 & ? & - \\
 WDc-8 & 23 21 14.38 & -13 27 36.8 & 19.23 & 18.42 & 18.00 &    - &    - &    - &     &       &   & - \\

\hline
\multicolumn{13}{|l|}{Notes - ? Indicates uncertainty due to a small measured motion and high uncertainties associated with these measurements (see text).}\\
\end{tabular}
\label{photom_table}
\end{table*}

\section{Followup Observations $\&$ Data Reduction}
\subsection{Ultracool dwarf candidates}
Second epoch images of candidate UCDs were taken with the InfraRed
Imager and Spectrograph, IRIS2 on the Anglo Australian Telescope (AAT)
during service observations on 2006 July 7 and 2006 December 8,
with $J$-, $H$- and $K_{s}$- band filters. The images were reduced using
the standard {\sc oracdr} package for IRIS2; this included de-biasing, dark
subtraction, background subtraction, removal of bad pixels and
mosaicing of jittered images.

Spectroscopic observations of one candidate (UCD-1; see
Table~\ref{photom_table}) were also obtained with IRIS2 on the AAT on
2006 September 8. The long slit mode was used with a 1 arcsec slit width
in the $J$- long and $H$- short grisms, covering wavelength ranges
$1.1 - 1.33$\,$\mu$m and $1.46 - 1.81$\,$\mu$m with a dispersion of
0.225\,nm/pixel and 0.341\,nm/pixel, respectively ($R\sim$\,2400).  A
total exposure time of 20 minutes in each band was obtained and the
target was nodded along the slit in an ``ABBA'' pattern with
individual exposure times of 300s. Standard dome flats and Xenon arcs
were taken at the end of the night and an F5V star was observed at a
similar airmass to the target to provide telluric correction. The
observing conditions were reasonable with an average seeing of
$0.8-1.2$ arcsec.

\begin{figure*}
\includegraphics[width=100mm, angle=90]{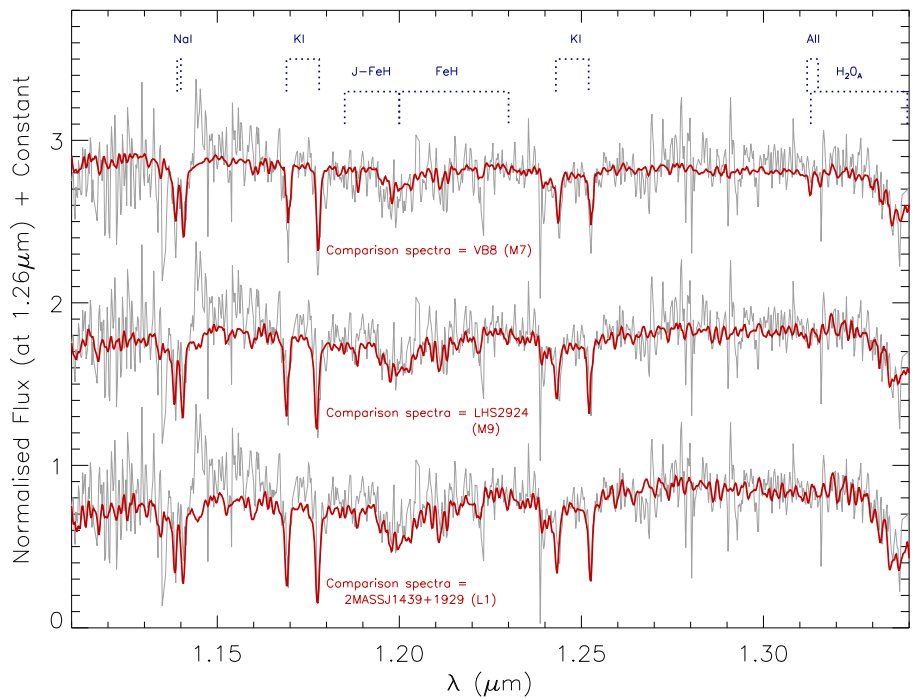}
\caption{$J$- band spectra of 2MASSJ$0030-3739$ (thin grey line), shown with
comparison spectra (thick line) of an M7, M9 and L1 type (top to
bottom) from \citet{cushing05} overlaid.}
\label{ucdspecj}
\end{figure*}

\begin{figure*}
\includegraphics[width=100mm, angle=90]{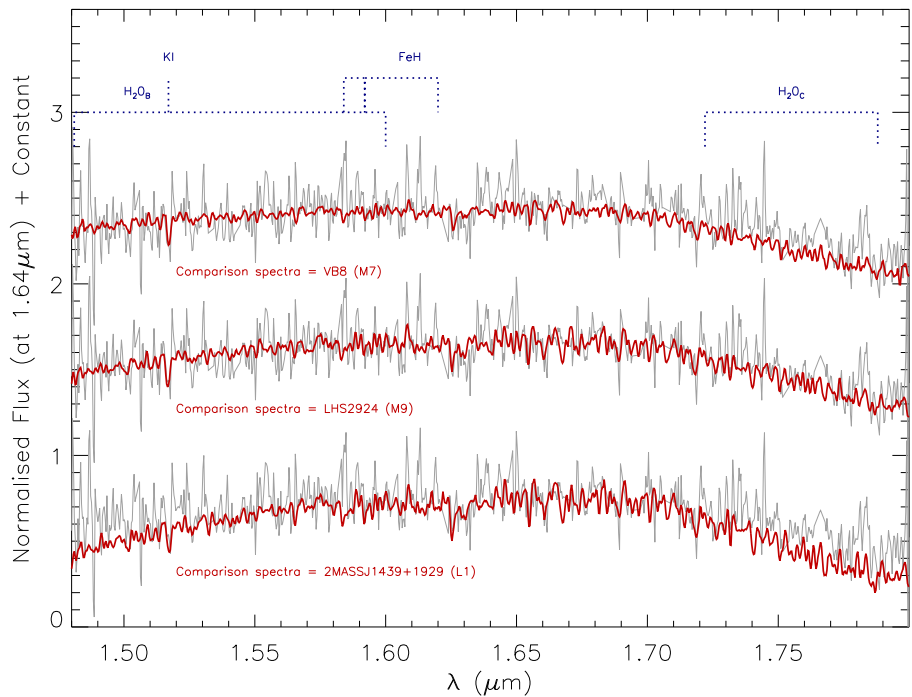}
\caption{$H$- band spectra of 2MASSJ$0030-3739$ (thin grey line). See
caption to Fig.~\ref{ucdspecj}.}
\label{ucdspech}
\end{figure*}

Standard {\sc iraf} routines were used to reduce the spectra including
flat fielding and cosmic ray removal. The spectra were extracted with {\sc
apall}, using a chebyshev function to fit the background and a third
order legendre function to trace the fit to the spectrum. The
wavelength calibration was done using the spectrum from a Xenon arc
lamp, using {\sc identify} to reference the arc lines and the {\sc
dispcor} routine to correct the dispersion of the spectrum. This
method was repeated for each of the differenced AB pairs, and the
wavelength calibrated spectra were median combined and flux calibrated
using the telluric standard and the spectra of a blackbody with a
$T_{\rm eff}$= 7500K. Annotated spectra are shown in
Fig.~\ref{ucdspecj} and Fig.~\ref{ucdspech}.

\subsection{White dwarf candidate}
A Spectrum of WDc-1 (2MASSJ$0030-3740$) was obtained with FORS1 on the
VLT on 2007 January 24, with directors discretionary time in program
278.C-5024(A), using the longslit mode in the optical wavelength range
$3800-5200$\,\AA\ and a dispersion of 50\,\AA/mm. Three integrations
of 600s were taken, giving a total exposure time of 30
minutes. Spectra of a DC WD and a standard F-type star were also taken
and used for calibration. Sky flats were taken and HgCd arcs were used
for wavelength calibration.

\begin{figure}
\vspace{3.5cm}
\hspace{2.0cm}
\includegraphics{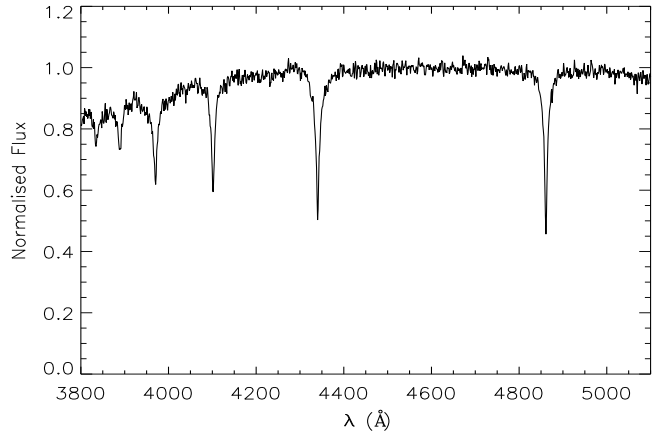}
\vspace{2.0cm}
\caption{Optical spectrum of the confirmed white dwarf WDc-1
(2MASSJ$0030-3740$), flux calibrated and normalised at 4600\AA.}
\label{wdspec}
\end{figure}

Standard IRAF packages were used to reduce the spectra including
debiasing, flat fielding and removal of bad pixels; the three spectra
were then extracted and wavelength calibrated as described in Section
3.1. The resulting spectra of both WDc-1 and the standard were divided
by the smooth DC WD spectrum, which has no intrinsic spectral
features, enabling correction for the instrumental response. The
standard star was then used for flux calibration and the final
spectrum of 2MASSJ$0030-3740$ is shown in Fig.~\ref{wdspec}.

\subsection{Proper motions}

The imaging data, as described in Section 3.1 were used to calculate
proper motions of our UCD candidates. The {\sc iraf} routines {\sc
geomap} and {\sc geoxytran} were used to transform between the
available multi-epoch images, using an average of 15 reference
stars. In addition a correction was applied to the derived proper
motions to account for the average (but small) proper motion of the
reference stars. This allowed any motion of the UCD candidates to be
accurately measured. Proper motion uncertainties were estimated from
centroiding accuracies combined with the residuals associated with our
derived transformations.

Amongst the eight candidate binary pairs, three of the UCD candidates
were ruled out since they are not common proper motion companions (at
$>$3\,$\sigma$), three remain uncertain due to the relatively small
motion expected between epochs (compared to the uncertainties
associated with the inter-epoch transforms) and one remains unobserved
(see Table~\ref{photom_table} for a summary of these results).

\begin{figure*}
\includegraphics[width=100mm, angle=90]{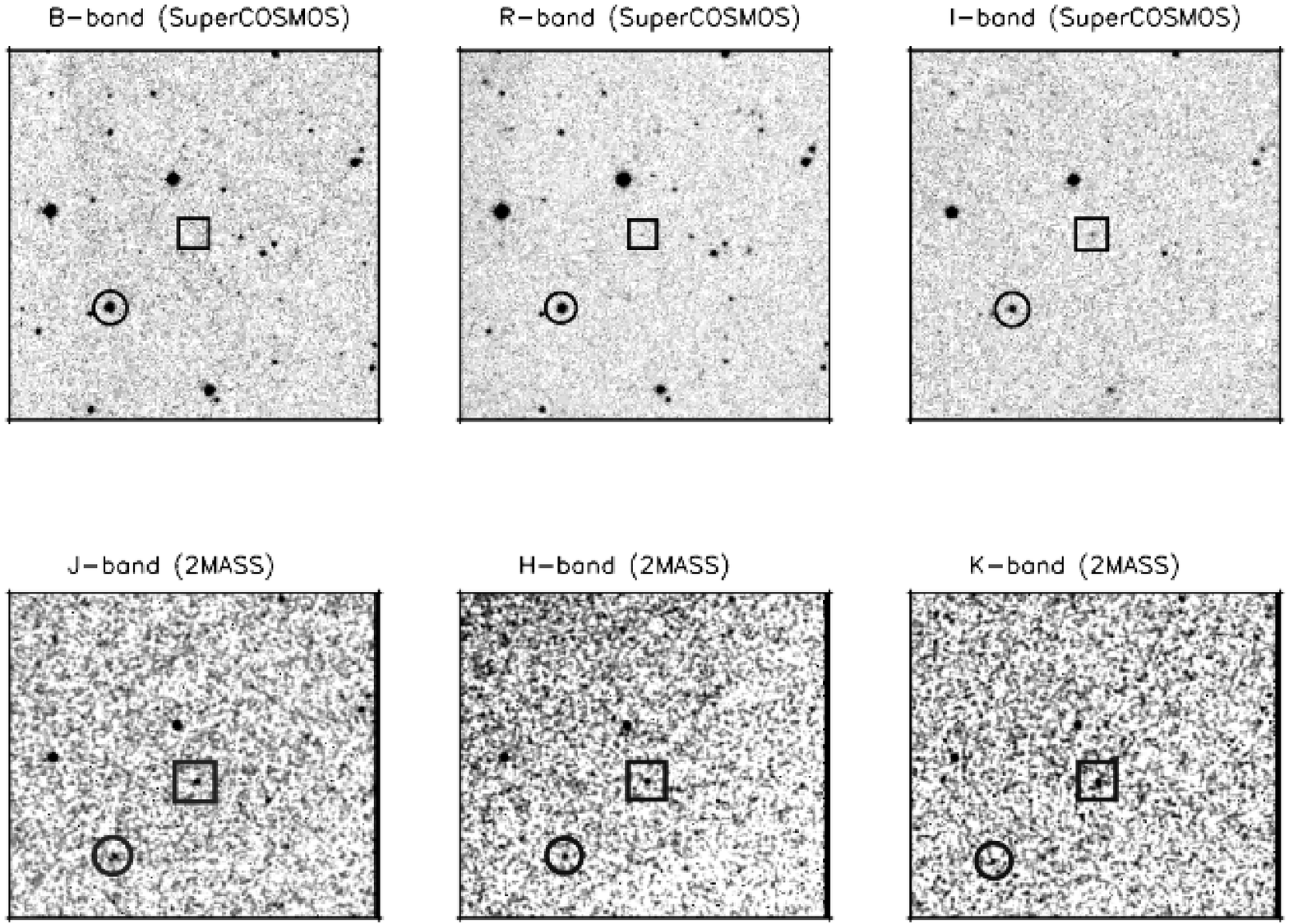}
\caption{SuperCOSMOS and 2MASS images show UCDc-1 (2MASSJ$0030-3739$;
squares) and WDc-1 (2MASSJ$0030-3740$; circles).}
\label{brijhk}
\end{figure*}

Second epoch IRIS2 measurements of candidate UCDc-1 (2MASS
J$0030-3739$) revealed a significant motion over the 6.8 year baseline
between the 2MASS first epoch and the IRIS2 second epoch images. The
large field of view of IRIS2 (7\,x\,7 arcmin) allowed us to measure the
proper motion of both the UCD and the WD candidate from the same
2-epoch image set, which clearly revealed the common proper
motion. Our final proper motion measurements were based on four
individual measurements: pairing up $J$-, $H$- and $K$- band images
from the 2MASS and IRIS2 epochs appropriately, as well as pairing up a
SuperCOSMOS $I$- band as first epoch with the IRIS2 $J$- band image as
second epoch. The last combination is over a relatively longer
baseline of 15.86 years, although may suffer from larger chromatic
effects due to the different bands. Our final proper motions were an
average of these four measurements, and the associated uncertainties
were estimated from their standard deviation.  Our measured proper
motions of the pair are given in Table~\ref{properties}.  Multi-band
$BRIJHK$ 2\,x\,2 arcmin finder charts centred on 2MASSJ$0030-3739$ are
shown in Fig.~\ref{brijhk}.

\section{Spectral Classification}
\subsection{UCDc-1: 2MASSJ0030-3739}
\subsubsection{Spectral ratios}
We estimate a spectral type for 2MASSJ$0030-3739$ based on spectral
ratios used in previously published work. Our $J$- band spectral
coverage is $1.1-1.35$\,$\mu$m. In this range we use the FeH ratio from
\citet{slesnick04}, the $J$- FeH and H$_{2}O_{A}$ ratio from
\citet{mclean03}. We measure a ratio of 0.854 for FeH, and using the
spectral type relation from \citet{slesnick04}, estimate a spectral
type of M9 from this ratio. The $J$- FeH ratio of 0.85 combined with
the spectral type relations from fig. 12 of \citet{mclean03} gives a
spectral range M8-L3. At the edge of our spectral coverage in the $J$-
band, when strong water vapour absorption starts to appear, we measure
the H$_{2}O_{A}$ ratio as 0.554. The relation between H$_{2}O_{A}$
and spectral type in fig. 11 of \citet{mclean03} indicates an $\sim$L1
type from this ratio.  \\

Our $H$- band spectrum covers wavelengths $1.45-1.81$\,$\mu$m, for which we used
H$_2$O ratios from \citet{reid01}[H$_2$O$_C$] and
\citet{mclean03}[H$_2$O$_B$]. The H$_2$O$_C$ ratio of 0.697 indicates
an L1.5\,$\pm$2 and the H$_2$O$_B$ ratio of 0.9 is consistent with an M8-M9
type. All the ratios we consider are indicated in Fig.~\ref{ucdspecj}
and Fig.~\ref{ucdspech}, and our spectral ratio results are summerised
in Table~\ref{spectype_table}.

\subsubsection{Comparison to template spectra}
Template spectra of known late M and early L dwarfs were used to make
a comparison to the overall profile of 2MASSJ$0030-3739$, as well as a
comparison to the depth of the absorption in spectral features such as
the KI (1.168, 1.179 and 1.243, 1.254\,$\mu$m), NaI(1.138,
1.141\,$\mu$m) and the AlI (1.311, 1.314\,$\mu$m) doublets in the $J$-
band; and the blended KI doublet (1.517\,$\mu$m) and FeH (1.58, 1.59
and 1.62\,$\mu$m) in the $H$- band. In general alkali metal lines
weaken at the M/L boundary \citep{mclean00}, but the NaI, AlI, FeH and
KI doublets are still clearly recognisable for the purposes of a
comparison. The spectra of an M7(VB8), M9(LHS2924) and
L1(2MASSJ$1439+1929$) from \citet{cushing05} were rebinned to the
dispersion of our near infrared spectra and normalised at 1.26 and
1.64\,$\mu$m in the $J$- and $H$- bands respectively
(Fig.~\ref{ucdspecj} and ~\ref{ucdspech}). Visual inspection of the
blended line features reveal the spectra are most consistent with an
$\sim$M9 type.

\subsubsection{Equivalent widths}

We calculated equivalent widths for the four neutral alkali (KI) lines
present in the $J$- band spectra using the methods of
\citet{mclean03}.  An {\sc idl} program was used to interactively
determine the equivalent width of each KI line, which were compared
with those of \citet{mclean03}, to estimate a spectral type for each
line (see their table 7). In order to reduce the amount of bias when
selecting the reference continuum, the process was repeated 12 times
using a continuum measured at different relative positions (within
0.05\,\AA\ from the centre of the line) and the mean of the
measurements used. A width of 4.89\,\AA\ at the 1.168\,$\mu$m line
indicates an M8/9; while the other three KI line widths at 1.177,
1.243 and 1.254\,$\mu$m are all consistent with an M7/8 type.

Analysis of the spectra of 2MASSJ$0030-3739$, through the use of
spectral ratios, comparison to template spectra and equivalent widths
(summerised in Table~\ref{spectype_table}) are all consistent with a
spectral type M9\,$\pm$1. We can use the relation between spectral
type and absolute magnitude from \citet{dahn02} to calculate a range
in $M_J$ of $10.85-12.04$ for the spectral range M8-L0. Thus,
combining this with the measured $J$- band magnitude from 2MASS, we
estimate 2MASSJ$0030-3739$ is at a distance of $41-75$\,pc.

\begin{table}
\caption{Estimated spectral types for 2MASSJ$0030-3739$.}
\centering
\begin{tabular}{|l|l|c|}
\hline
Method & Reference  & Spectral type \\
\hline
Ratio FeH (1.200/1.230\,$\mu$m) & Slesnick ('04) & M9 \\
 $''$  $J$- FeH (1.185/1.200\,$\mu$m) & McLean ('03) & M8-L3 \\
 $''$   H$_{2}O_{A}$ (1.313/1.343\,$\mu$m) &  $''$          & $\sim$L1 \\
 $''$   H$_{2}O_{C}$ (1.722/1.788\,$\mu$m) & Reid ('01) & L1.5\,$\pm$2 \\
 $''$   H$_{2}O_{B}$ (1.480/1.600\,$\mu$m) & $''$ & M8-M9 \\
SC $J$- & ... & M9\,$\pm$1 \\
SC $H$- & ... & M9\,$\pm$1 \\
EW KI (@ 1.168\,$\mu$m) & McLean ('03) & M8-M9 \\
EW KI (@ 1.177\,$\mu$m) & $''$ & M7-M8 \\
EW KI (@ 1.243\,$\mu$m) & $''$ & M7-M8 \\
EW KI (@ 1.254\,$\mu$m) & $''$ & M7-M8 \\
\hline
\multicolumn{3}{|l|}{Notes: SC- Spectral Comparison, EW- Equivalent Width.}\\
\end{tabular}
\label{spectype_table}
\end{table}

\subsection{WDc-1: 2MASSJ0030-3740}

\begin{figure}
\vspace{4.0cm}
\hspace{2.0cm}
\includegraphics{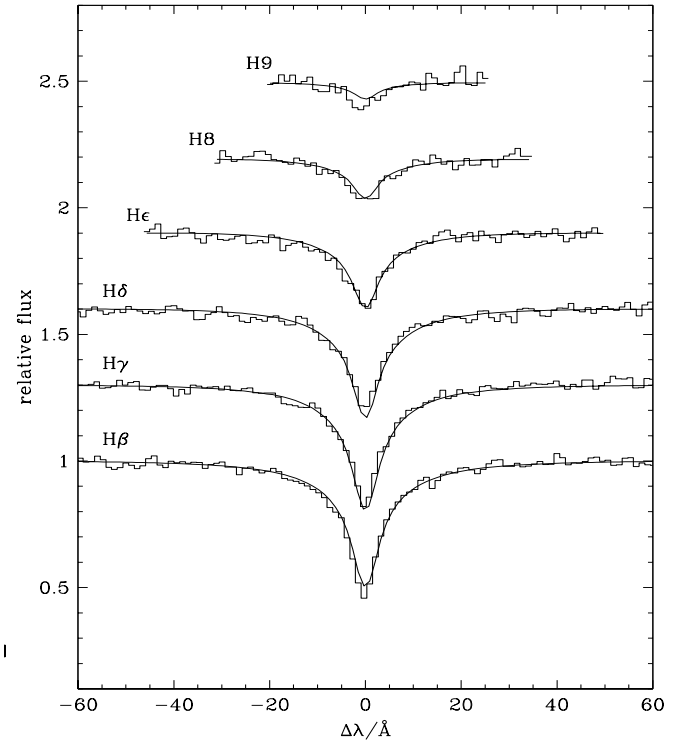}
\vspace{6.0cm}
\caption{Model atmosphere fit of the Balmer lines of 2MASSJ$0030-3740$ with our
best fit parameters $T_{\rm eff} = 7600$\,K and log
($g$)$=7.79$. Observed and theoretical fluxes were normalised to the
continuum. Line profiles are shifted for clarity.}
\label{wdfit}
\end{figure}

We derived WD parameters, $T_{\rm eff}$ and log($g$)
from a fit of the Balmer lines using the fitting routine {\sc FITPROF}
described in \citet{napiwotzki99}.  The WD spectrum is
modelled using an extensive grid of spectra computed with the model
atmosphere code of Detlev Koester described in \citet{finley97}.
Observational and theoretical Balmer line profiles are normalised to a
linear continuum and the atmospheric parameters determined with a
$\chi^2$ algorithm. Our best fit is shown in Fig.~\ref{wdfit}. Results
are $T_{\rm eff}=7600\pm 20$\,K and log($g$) = 8.09\,$\pm 0.04$
(formal errors from the fit routine to one sigma).

Realistic error estimates are often substantially larger than the
formal statistical estimates. Napiwotzki et al. (1999) derived a relative
uncertainty of 2.3 per cent in $T_{\rm eff}$ and 0.075\,dex in log($g$) from a
sample of DA WDs analysed using the same fitting method.  We
adopt this temperature uncertainty for our following estimates, but the
gravity determination is likely subject to a systematic overestimate.
The mass distribution of WDs peaks close to $0.6$\,M$_\odot$
(e.g. Napiwotzki et al. 1999; \citealt{bergeron07}), corresponding to
log ($g$) values close to 8.0. Spectroscopic investigations applying
the method explained above show a trend of the log($g$) distribution
peaking at increasingly higher values for decreasing temperatures
(Bergeron et al.\,2007).  This trend starts at 11500\,K putting the white
dwarf analysed here in the affected region.  Taken at face value, this
would indicate on average higher masses for cool WDs.
However, as argued by Bergeron et al.\,(2007) and \citet{engelbrecht07}
this can be ruled out. A source of extra line broadening must be
present in these stars. However, the exact nature of this mechanism is
still under discussion and as pointed out in Bergeron et al.\,(2007)
anecdotal evidence suggests that some stars are affected and others
are not.

Here we take a pragmatic approach. We used a large sample of DA white
dwarfs observed in the course of the SPY programme
\citep{napiwotzki01} to derive an empirical correction.  Spectra of
these stars were analysed using the same model grid and very similar
analysis methods as the WD discussed here. We estimated a
shift of the log($g$) distribution caused by this unknown mechanisms
of 0.3\,dex in the $T_{\rm eff} = 7500 - 8000$\,K range. The
corrected gravity is then log($g$)=7.79. As mentioned above it is not
entirely clear whether all cool WDs are affected. Thus we
will use the corrected and uncorrected gravity values for further
discussion. WD masses and cooling ages were calculated by
interpolation in the \citet{benvenuto99} cooling tracks for white
dwarfs with thick hydrogen envelopes. Results are listed in
Table~\ref{wd_properties}.

\begin{table}
\caption{Fit results and derived quantities of WD mass, cooling age
and absolute brightness for the corrected and uncorrected case
discussed in Section 4.2.}
\begin{tabular}{llllll}\hline
     &$T_{\mathrm{eff}}$ &log ($g$) &M &$t_{\mathrm{cool}}$ &$M_V$\\
     &[K] & [$\rm {cm\,s^{-2}}$] & [M$_\odot$]&[Gyr] & \\ \hline
     solution 1 &7600 &8.09 &0.65 &1.48 &13.4 \\ solution 2 &7600
     &7.79 &0.48 &0.94 &12.9 \\ \hline
\end{tabular}
\label{wd_properties}
\end{table}

We used the latest table of DA colours from \citet{holberg06} combined
with the model $M_V$ values from Table~\ref{wd_properties}, to
calculate a range in $M_I$ for $T_{\rm eff}$ ($7600$\,$\pm175$\,K) and
log($g$) estimates. Combined with the well calibrated $I$- band
magnitude from DENIS we determined a distance for 2MASSJ$0030-3740$ of
$27-59$\,pc, which is consistent with our distance estimate for the
UCD companion 2MASSJ$0030-3739$. We thus suggest our binary is at a
distance of $41-59$\,pc, which gives a binary separation of
$3650-5250$\,AU.

\section{Discussion}
\subsection{A randomly aligned pair?}

In order to determine if the new system is a bonafide UCD-WD binary,
we have statistically assessed the likelihood that two such objects
could be a line-of-sight association with photometry and proper motion
consistent with binarity by random chance. To do this we began with
the UCD luminosity function of Cruz et al. (2007), which gives a
number density of (4.9\,$\pm$0.6)$\times 10^{-3}$ UCDs per $\rm
{pc^{-3}}$. We then calculated a volume associated with 1532 circular
areas on the sky (one for each of our WD candidate sample), with radii
of 89 arcsec (separation of the components) and a line-of-sight depth
of 58\,$\pm$17pc (approximate distance to the new M9 UCD, using
relations from \citealt{dahn02}).  This volume equates to 58\,$\pm$26
$\rm {pc^{-3}}$, giving a total expected number of 0.28 UCDs to be
within 89 arcsec of one of our WD candidates.

To factor in the probability that two objects might have a common
proper motion at the level as our measurements, we downloaded a
magnitude-limited sample ($R$\,$<$20) from the SuperCOSMOS Science
Archive, applying the same minimum proper motion requirement that was
used to create our WD candidate sample. This sample of 160 sources was
centred on our WD and selected from a large circular sky area of
radius 90 arcmin. We then constructed a proper motion
vector-point-diagram, and counted sources that were found to be within
the 2\,$\sigma$ uncertainty circle of the measured UCD proper motion. We
found that four of the 160 sources had proper motion consistent with
our UCD, suggesting a probability of 2.4\,$\pm$1.2 per cent that such common
proper motion could occur by random chance (where we assume Poisson
uncertainties associated with this and other samples considered in
this discussion).

An additional requirement fulfilled by our UCD-WD system is that the
colour-magnitude information must be consistent with a common distance
(see Fig.~\ref{ld_cmd}), and we found that 53 per cent ($786\,\pm28$) of our WD
candidate sample were photometrically consistent with being at the
same distance as the UCD.

Finally we consider what fraction of our WD candidates might
be spurious, and thus not able to contribute to non binary
line-of-sight associations where the WD has been confirmed
spectroscopically. In the magnitude range $R$\,$<14$, where MS99 is
thought to be essentially complete, we find that 66 per cent of our WD
candidates are included in the MS99 catalogue. This suggests that, at
least for brighter magnitudes, our WD candidates are relatively free
from contaminating objects, and that our selection techniques are
robust. While we cannot be sure that the same low-level of
contamination applies to the full magnitude range, we take the
conservative approach and consider the full WD candidate sample as
potentially contributing to non-binaries that appear to be UCD-WD
pairs.

Taking into account all these factors, we estimate that we would
expect $0.0036\,\pm0.0025$ randomly aligned UCD-WD pairs with
$\le$\,89 arcsec separation and proper motion and photometry
consistent with binarity at the level of our observations. The
likelihood of the system being merely a line-of-sight association is
thus vanishingly small, and we can assume that the UCD-WD pair is a
gravitationally bound binary system.

\subsection{Binary age}

The age of the binary system can be constrained from the white dwarf
mass and cooling age. For our corrected fit, we suggest that
2MASSJ$0030-3740$ is 0.48\,M$_\odot$ and has a cooling age of
0.94\,Gyr. We access the IFMR determinations of \citet{weidemann00},
\citet{dobbie04}, \citet{dobbie06}, \citet{ferrario05},
\citet{catalan08} and \citet{kalirai08} to estimate a likely,
initial-mass constraint for the main-sequence progenitor star of
$1-2$\,M$_\odot$.  The main sequence lifetime of a progenitor of this
mass is likely $>$1Gyrs (probably several Gyrs) and thus is not useful
when trying to constrain the upper age limit of the system. Thus the
upper age limit of our system remains uncertain. Note that if the WD
mass were higher, for example if the helium enrichment of the
atmosphere is lower than typical (see Section 4.2) then the WD's
log($g$) could be as high as 8.1\,dex, with a WD mass of
$\sim$0.65\,M$_\odot$ and cooling age of $\sim$1.5\,Gyrs. This would
allow progenitor mass to be constrained to a likely range of
$<2.7$\,M$_\odot$, giving a main-sequence progenitor lifetime of
$<0.83$\,Gyrs \citep{monteiro06}, yielding a binary age constraint of
$1.5-2.3$\,Gyrs. This possibility is instructive at least, in
demonstrating the level of age constraints (with accompanying UCD
constraints) that may be placed on benchmark binaries of this
type. However, it is not possible to judge the helium content of the
WD's atmosphere (if any), and we can only confidently place a lower
limit on the age of this binary from our best fit cooling age for the
WD and the likely length of the main sequence lifetime of the
progenitor, which is likely equal to or larger than the white dwarf
cooling age. The age of the binary is thus $>1.94$\,Gyrs.

\subsection{UCD properties}

We have estimated $T_{\rm eff}$, mass and log($g$) from the Lyon group
dusty models (\citealt{chabrier001}; \citealt{baraffe02}), using the
minimum age of the system (0.94\,Gyrs) and our estimated $M_J$ of
2MASSJ$0030-3739$. The models indicate that 2MASSJ$0030-3739$ has
$T_{\rm eff}=2000-2400$\,K, mass=$0.07-0.08$\,M$_{\odot}$ and
log($g$)=$5.30-5.35$, placing it close to the limit for hydrogen
burning. Note that our $T_{\rm eff}$ is consistent with the
semi-empirical estimates of \citet{golimowski04} for an M9\,$\pm$1
dwarf, which use well measured luminosities and a model constraint on
radius (which changes by $<$10 per cent for ages of $1-5$\,Gyrs), to
determine $T_{\rm eff}$ values spanning a wide range of spectral
type. The full list of properties for the binary are listed in
Table~\ref{properties}.

\begin{table}
\caption{Parameters of the binary and it's components.} \centering
\begin{tabular}{|l|l|c|}
\hline
Parameter &  & Value \\
\hline
Separation on sky & ............ & 89 arcsec\\
Estimated distance & ............ & $41-59$\,pc\\
Estimated line-of-sight \\
separation & ............ & $3650-5250$\,AU\\
Minimum age of system & ............& $>1.94$ \,Gyrs\\
& & \\
\underline{Ultracool Dwarf} &  &  \\
RA & ............ & 00 30 06.26\\
DEC & ............ & -37 39 48.2\\
2MASS designation & ............ & 2MASSJ$0030-3739$\\
Distance & ............ & 41 -- 75pc\\
2MASS $J$  & ............ & 15.2\,$\pm$ 0.05\\
2MASS $H$  & ............ & 14.4\,$\pm$ 0.05\\
2MASS $K_s$  & ............ & 13.8\,$\pm$ 0.06\\
DENNIS $I$  & ............ & 18.4\,$\pm$ 0.23\\
DENNIS $J$  & ............ & 15.06\,$\pm$ 0.14\\
SuperCOSMOS $I$  & ............ & $\sim$18.3\\
$\mu$ RA  & ............ & -130\,$\pm$30\,mas yr$^{-1}$\\
$\mu$ DEC  & ............ & -70\,$\pm$20\,mas yr$^{-1}$\\
Spectral Type  & ............ & M9\,$\pm$1\\
Mass & ............ & $0.07-0.08$\,M$_\odot$\\
$T_{\rm eff}$ & ............ & $2000-2400$\,K\\
log($g$) & ............  & $5.30-5.35$\,dex \\
 & & \\
\underline{White Dwarf} &  &  \\
RA & ............ & 00 30 11.9\\
DEC & ............ & -37 40 47.2\\
2MASS designation & ............ & 2MASSJ$0030-3740$\\
Distance & ............ & $27-59$\,pc\\
2MASS $J$  & ............ & 16.1\,$\pm$ 0.11\\
2MASS $H$  & ............ & 15.8\,$\pm$ 0.15\\
DENNIS $I$  & ............ & 16.2\,$\pm$ 0.07\\
DENNIS $J$  & ............ & 15.9\,$\pm$ 0.22\\
SuperCOSMOS $B$  & ............ & $\sim$16.77\\
SuperCOSMOS $R$  & ............ & $\sim$16.35\\
SuperCOSMOS $I$  & ............ & $\sim$15.97\\
$\mu$ RA  & ............ & -83\,$\pm$30\,mas yr$^{-1}$\\
$\mu$ DEC  & ............ & -70\,$\pm$12\,mas yr$^{-1}$\\
Spectral Type & ............& DA\\
$T_{\rm eff}$  & ............ & 7600$\pm$175\,K\\
log($g$)  & ............ & $7.79-8.09$\,dex\\
Mass  & ............ & $0.48-0.65$\,M$_{\odot}$\\
WD cooling age & ............ & $0.94 - 1.5$\,Gyrs\\
WD progenitor age & ........... & $>1$\,Gyr\\

\hline
\end{tabular}
\label{properties}
\end{table}

\section{Future Work}
A parallax measured distance would allow the luminosity of
2MASSJ$0030-3739$ to be directly measured. Our age constraint would
then facilitate a tight model constraint on the radius, and thus an
accurate determination of $T_{\rm eff}$ and log($g$). This UCD could
then have a very useful role as part of a testbed for our
understanding of ultracool atmospheres. For example,
\citet{mcgovern04} studied a set of M dwarfs with young and
intermediate ages, and showed that a variety of spectral lines
including the $J$- band KI lines, are gravity sensitive. It is also
clear that the effects of dust in very late M dwarf atmospheres are
not well understood (e.g. \citealt{jones05}). Removing degeneracies in
the physical properties of such objects will allow for a more focused
testing of available models in this region of $T_{\rm eff}-$log($g$)
space.

A more expansive search for benchmark UCD-WD binary systems can be
made using the combination of UKIDSS and SDSS (e.g.
\citealt{day-jones07}). These surveys probe a significantly larger
volume for UCDs and WDs than 2MASS and SuperCOSMOS, and will thus
yield a larger number of wide benchmark binaries, with a broader range
of properties.  It is likely that significant numbers of binary
systems could be identified containing an L/T dwarf and a WD (see
\citealt{pinfield06}). Also, if the frequency of wide UCD companions to
higher mass stars is of a similar level to those found around solar type
stars, there could be numerous detectable systems with high mass WD
components. Tight constraints on binary age will be particularly
important when determining the properties of cooler, lower mass UCD
companions, since sub-stellar brown dwarfs cool continuously, and
their mass and surface gravity depend strongly on age. The broad band
spectral morphology of L and T dwarf populations show a significant
scatter independent of spectral type, presumably due to variations in
surface gravity and metallicity. Thus, the physical constraints
offered by benchmark systems of this type could hold great potential
to reveal trends linking such spectral changes to variations in UCD
physical properties.

\section*{Acknowledgments}
We acknowledge Science and technology Facilities Council (STFC) for
their support to ACD, BB, SLF, DJW and JRC, we would also like to
acknowledge the helpful advice given from Detlev Koester in this
work. This publication has made use of the NASA/IPAC Infrared Science
Archive, which is operated by the Jet Propulsion Laboratory,
California Institute of Technology, under contract with the National
Aeronautics and Space Administration. We have also made use of the
data obtained from the SuperCOSMOS Science Archive, prepared and
hosted by the Wide Field Astronomy Unit, Institute for Astronomy,
University of Edinburgh, which is funded by the STFC. We gratefully
acknowledge the UK and Australian government support of the
Anglo-Australian Telescope through their STFC and DETYA funding as
well as NASA grant NAG5-8299 \& NSF grants AST95-20443 and AST-9988087
and Sun Microsystems. We would also like to acknowledge the DENIS
project, that has been partly funded by the SCIENCE and the HCM plans
of the European Commission under grants CT920791 and CT940627. It is
supported by INSU,MEN and CNRS in France, by the State of
Baden-W\"urttemberg in Germany, by DGICYT in Spain, by CNR in Italy,
by FFwFBWF in Austria, by FAPESP in Brazil, by OTKA grants F-4239 and
F-013990 in Hungary, and by the ESO C\&EE grant A-04-046. Jean Claude
renault from the IAP was the project manager. Observations were
carried out thanks to the contribution of numerous students and young
scientists from all involved institutions, under the supervision of
P. Fouqu\'e, survey astronomer resident in Chile.

\bsp

\label{lastpage}

\end{document}